# MODELING THE THERMO-MECHANICAL BEHAVIOR OF A WOVEN CERAMIC MATRIX COMPOSITE AT HIGH TEMPERATURES

Koffi Enakoutsa[1,3], Youssef Hammi[2], John E. Crawford[1],
Joseph Abraham[1], and Joe Magallanes[1]

[1]*Karagozian & Case, Inc., 700 N Brand Boulevard Suite # 700, Glendale, CA, 91203;*
[2]*Mississippi State University Center for Advanced Vehicular Systems, 200 Research Boulevard, Mississippi State, MS, 39759*
[3] *International Research Center on Mathematics and Mechanics of Complex Systems, Universita dell'Aquila, Cisterna di Latina, Italy*

**ABSTRACT**

Some years ago, (Ladeveze, P 1983) has proposed a continuum damage model which is based on a sound thermodynamic theory and has the potential to capture the physics of ceramic matrix composite (CMC) damage mechanisms, including matrix cracking and multi-axial fiber degradation. These mechanisms are introduced in the model *via* an anisotropic damage formulation that accounts for micro-crack opening and closure effects. This paper aimed at extending (Ladeveze, P 1983)'s model to capture thermal expansion and coupled thermo-mechanical phenomena, as such as those encountered at elevated and extreme temperatures. In the new model a linear thermal expansion coefficient is added to equation of state to account for the thermal to mechanical coupling effects. The mechanical to thermal effects are introduced by assuming an internal heat generation due to residual strain effects. The constitutive relations of the coupled thermo-mechanical behaviors of CMC material model were incorporated into *ABAQUS©* finite element code as *UMAT* software. The numerical algorithm bears some resemblance to (Genet, M; Marcin, L.; Ladeveze, P 2013)'s algorithm, which essentially consists of a local loop made of nested fixed-point and Newton-Raphson iterations, the former to calculate the damage state and the latter to invert a nonlinear state law, but departs from it in the numerical implementation of the residual strain part of the total strain where a radial return algorithm was adopted. An extension of the methodology presented by (Letombes, S. 2005) to calibrate the CMC damage model parameters based on a woven silicon-carbide/silicon-carbide (*SiC/SiC*) material developed by SAFRAN Group for which experimental data are available in the open-literature will be provided. Feasibility tests (tension-compression, tension-tension cyclic, simple tension tests) based on a dog bone and open hole specimen tests demonstrated a good capturing of nonlinear as well as coupled thermos-mechanical behaviors in the *SiC/SiC* material. Future works will extend the CMC material model to include creep, fatigue, and coupling chemical and mechanical effects which are inherent to high temperature deformation and degradation of CMCs.

## 1.0 INTRODUCTION

Current challenges in aeronautical industries include the need to realize reduction in fuel consumption and/or increases in the aircraft performance in extreme environments. The use of high performance composite materials, such as ceramic matrix composites (CMCs), affords a means to meet these challenges. Indeed, CMC materials have a lower density, offer an excellent resistance to thermo-mechanical fatigue loadings, and are particularly valuable for gas turbine hot section components and hypersonic leading edge materials. CMC materials have a longer service life and thereby constitute an excellent alternative to metallic alloys in several aeronautical applications: combustion chambers, turbines shrouds, gas turbines, and nozzles where sharp temperature gradients are present.

### 1.1.1 NEED FOR COMPREHENSIVE CMC MATERIAL MODELS

To effectively support the use of CMCs requires that analytic methods (i.e., material models) be developed to compute their thermal, mechanical, and coupled thermo-mechanical behavior/responses. These models include considerations pertaining to computing responses under extreme environments that induce severe thermal gradients and that must be able to capture degradation and failure of CMCs in such environments. Experimental approaches to understand CMC material inelastic behavior are very expensive. Also the study of lower level systems failures resulting from extreme external conditions, such as those encountered, for instance during hypersonic flights, propulsion stages, for large scale complex systems is a primary research area that requires additional investigations. Alternatively, the use of validated predictive computational physics-based models which can describe accurately the different stages of CMCs' behavior until failure are preferred. Such a predictive tool could enable the virtual study of structural behavior under extreme conditions and allows parametric studies to be performed to evaluate opportunities to prevent and mitigate these issues. Current finite element (FE) models used to perform analyses of structures composed of CMCs are often limited in scope to linear elastic regime which fails to capture the strain and stress localization zones which are precursors of damage and fracture in a material under extreme thermo-mechanical loadings. Therefore, there is a need to develop a predictive tool that accounts for the progressive accumulation of such damage and effectively assess whether the influence this localized damage on the overall performance of the system within which it is a part which inevitably leads to material failure when the latter is subjected to extreme conditions.

### 1.1.2 CRACKING MECHANISMS IN CMCS

The initiation of mechanical degradation in CMCs is attributed to the formation of various networks of matrix cracks; a hierarchy of such networks is presented in (Forio and Lamon 2001). Several damage mechanisms could occur: matrix micro-cracking, fiber/matrix debonding, and fiber breakage. These mechanisms are strongly anisotropic: cracks could be normal to the loading direction or partly deviated by the reinforcement orientation. Furthermore, these cracks could be opened or closed depending on the loading and on the thermal residual stresses induced by cure processing. More precisely, under tensile loading, CMCs present a linear elastic response until the initiation and propagation of matrix micro-cracks and the partial re-opening of thermal cracks. These cracks mainly initiate at the singularity of macro-pores and propagate normally to the applied load direction. In a second stage, multiplication of matrix micro-cracks and the associated fiber/matrix debonding are propagating until matrix crack saturation, see (Guillaumat 1994). Composite with a weak interface exhibits a "plateau-like behavior", (Naslain 1993). The matrix crack saturation is rapidly achieved (load transfer being poor) and the total failure occurs almost immediately after the saturation point. For composites that present high strain to rupture, after matrix saturation, a significant domain related to a progressive load transfer to the fibers, which then fracture progressively, has been observed (El Bouazzaoui, Baste and Camus 1996). Another set of cracks corresponding to multiple cracking of the bundle may also occur, (Bale, et al. 2012).These composites represent a broad non-linear domain and allow for higher stresses without any plateau-like domain.

### 1.1.3 FAILURE MODELS

A number of micromechanics analyses have been developed for predicting the onset and progression of failure within brittle composites (Curtin 1993), (Lee and Daniel 1992), (Weitsman and Zhu 1993), and (Hedgepeth 1961). In what is now considered a classical analysis, (Aveston, Cooper and Kelley 1971) discussed in detail the "energetics of multiple fractures" in brittle composites, see (Curtin 1993). This work has fueled similar studies over several years. Many of the models presented over this time period are based upon the classical shear-lag formulation presented by (Hedgepeth 1961). The approach parallels the method employed by (Cox 1952) who first investigated the influence of a single short fiber embedded in an infinite medium, (Carrere and Lamon 1999), but can be adapted to investigate the response of unidirectional and cross-ply laminates if an equivalent damage state for the laminate can be determined. Unfortunately, this can be quite a difficult task. The complexities of brittle failure in composite materials have forced many researchers to rely on empirical data which has thereby reduced the utility of the analytical models. In addition, many existing analytical solutions employ failure criteria which significantly over-predict the rate of matrix cracking. The most obvious case is the original ACK-model in which all of the matrix cracks were assumed to form at a single applied stress (Curtin 1993). This resulted in a "stepped" or "plateaued" stress-strain response where the material response curve is initially linear followed by a single large jump in strain during matrix failure then the response becomes linear again albeit with a smaller slope. Hence, even though the micromechanics approach is appealing because of its simplicity, solutions from many of the existing models do not mirror experimental data, (Weitsman and Zhu 1993) and (Evans, Domerque and E. 1994) and, therefore, alternate approaches are sought.

Within the literature, there are a number of more detailed analyses which avoid some of the simplifying assumptions employed under the micromechanics approach, (Nairn 1990) and (Larson 1992). For example, a number of solutions employ traditional fracture mechanics techniques to investigate the conditions for crack growth near a bi-material (fiber/matrix) interface, (Chawla 1987) and (Han, Hahn and Croman R.B 1988). These models are useful since the development of valid design and failure criteria are contingent upon a full understanding of the micro-structural behavior of the laminate during loading. Unfortunately, modeling the behavior of an individual crack in this manner may require integration of many complex theories, e.g. linear elastic fracture mechanics, statistical analysis and variational mechanics; therefore, when considering the large number of cracks which are continually developing and growing in a CMC, the analysis can be quite complex. To further compound the problem, the crack formation within the composite is dependent not only on the lamina properties, but also on laminate and component geometries. In addition, matrix cracking is not the only damage mode observed in CMCs. Cracks can also develop within the fibers or along the fiber/matrix interface. Since the evolution of all these types of damage is dependent on the magnitude and type of loading, the operating environment must also be accounted for in the analysis considering all these effects, the problem quickly becomes overwhelming. Perhaps this explains why a large number of first-order models have been reported in the literature, (Pryce and Smith 1992) and (Spearing and Zok 1993).

In addition, different macroscopic damage models based on the influence of the major microscopic mechanisms on the macroscopic behavior of the material were also proposed. In these models, the crack closing/opening phenomena can be accounted for by decomposing the tensile/compression elastic density energy, (Ladeveze, P 1983) and (Hild, Burr and Leckie 1996). The crack micro-mechanisms of satin based CMCs manifest themselves at the macroscopic scale by some complex anisotropic damage. Indeed, each network of cracks induced a specific evolution law for the damage. Some of the models, see (Evans, A.G; Marshall, D.B 1989), were developed based on microscopic models. However, one difficulty lays in the description of the damage for complex loadings or outside of the fibers' axis, since the micro-models used only represent the behavior in the direction of the fibers. In fact, one of the modeling major difficulty is to account for the complex cracks' network because it is oriented by either the direction of the loading or that of the fibers. Models the kinematics of the damage of which is fixed by a referential related to the composite allows an easy description of the cracks networks oriented by the fibers, (Camus 2000), (Gasser, A. 1993), (Gasser, A.; Ladeveze, P; Peres, P. 1998), and (Gasser, A.; Ladevèze, P; Poss, M 1996), while models described in a referential related to the loading enable a better description of the whole network, (Rouby and Reynaud 1993).

Mechanical tests under cyclic loadings have evidenced a fatigue phenomenon which manifests itself by an evolution of the damage, a residual deformation and an evolution of the hysteresis loops of loading and unloading. An explanation of this phenomenon was given in (Evans, A.G 1997) and (Burr, Hild and Leckie 1998)based on a ruin mechanism of the interfacial fibers/matrix debonding in the direction of the fibers. These authors show that the fatigue damage of CMCs is partially driven by mechanical cycling. In addition, the ruin mechanism allows explaining the difference of the fatigue behavior of some CMCs in the presence of temperature. A description of the ruin mechanism at the interface fibers/matrix was used in (Bodet, et al. 1995) to define a fatigue damage evolution law for a macroscopic model.

### 1.1.4 NEW FORM OF CMC MATERIAL MODEL

An outline for the development of a new form of CMC material model is described in this paper. The intent is to obtain a formulation for a robust methodology for predicting the thermo-mechanical behaviors of these materials, particularly related to the cracking of the matrix (i.e., fracture responses) that might result in aerospace applications.

The CMC material model proposed herein relies on the classical modeling framework of the thermodynamics of irreversible processes with internal state variables, (Letombes, S. 2005) and (Baranger, E.; Cluzel, C.; Ladeveze, P.; Mouret, A. 2007). The different steps of the formulation are: (1) selection of the state variables and of a potential (i.e., the interpretation of these variables is given by a potential form); (2) computation of the associated thermo-dynamical forces by duality with state variables; (3) computation of the state laws; and (4) definition of the damage evolution laws and "ad hoc" failure criterion. The proposed model formulation takes into account the crack networks and the associated fiber-matrix debonding through damage and inelastic strains. The inelastic part is relatively classical and the damage part is based on anisotropic and unilateral damage theory, a powerful approach introduced in (Ladeveze, P 1983) and (Hild, Burr and Leckie 1996) which was applied to *SiC/SiC* composites (Camus 2000), (Gasser, A.; Ladevèze, P; Poss, M 1996) and concrete (Baranger, E.; Cluzel, C.; Ladeveze, P.; Mouret, A. 2007). An extension of (Ladeveze, P 1983) formulation for CMCs to account for thermo-mechanical coupling effects is introduced in a way that the stress analysis depends on the temperature distribution and the temperature distribution depends on the stress solution.

A numerical implementation of the newly obtained CMC model which bears some resemblance to (Genet, M.; Marcin, L.; Baranger, E.; Cluzel, C.; Ladeveze, P.; Mouret, A . 2011), (Cluzel, et al. 2009) and (Genet, M; Marcin, L.; Ladeveze, P 2013)'s numerical algorithm is proposed and incorporated into ABAQUS© FE software as a UMAT. Just like in (Genet, M; Marcin, L.; Ladeveze, P 2013), the numerical algorithm consists of local loop made of nested fixed-point iterations and Newton-Raphson iterations, the former to compute the damage state and the latter to invert the state which is non-linear, even when all the state variable are fixed. However, the implementation herein departs from Genet et al.'s algorithm in the numerical implementation of the residual strain part of the total strain where a radial return-like algorithm was applied.

An extension of a methodology presented in (Letombes, S. 2005) is used to calibrate the parameters of the CMC material model. The calculation is based on a *SiC/SiC* material developed by SAFRAN Group for which experimental data are available in the open literature. Simplified material response (i.e., for tension-compression, tension-tension cyclic, simple tension tests) are shown to demonstrate the capability of the CMC material model. They are based on dog bone and open hole specimen tests, which demonstrated a good capturing of the non-linear material behavior in the *SiC/SiC* material. The remaining part of the paper is organized as follows.

- Section 2 provides the constitutive relations for the coupled thermo-mechanical behavior of the CMC material. This model consists of the extension of (Ladeveze, P 1983) and co-workers model, which is viewed in Appendix A for CMC material, including temperature effects.

- Section 3 gives a detail description of the numerical algorithm used to incorporate the constitutive relations of the CMC model into ABAQUS FE code.

- Next, Section 4 describes a procedure based on (Letombes, S. 2005)'s work to identify the CMC material model parameter.

- Finally, Section 5 presents the verification and the validation process, conducted for the proposed CMC damage model.

## 2.0 CMC MATERIAL MODEL

As mentioned, the CMC model is an extension of one previously developed by (Ladeveze, P 1983) to incorporate thermal expansion and coupled thermo-mechanical effects. In Ladeveze's model, damage effects are coupled with the inelastic behavior of the CMC and thermal and coupled thermal and mechanical effects were disregarded. A detailed description of the procedure leading to the constitutive relations of Ladeveze et al.'s model can be found in (Cluzel, et al. 2009), (Baranger, E.; Cluzel, C.; Ladeveze, P.; Mouret, A 2007), (Letombes, S. 2005) etc. A review of these relations is given in the Appendix A to fully grasp the physics of damage in CMCs.

### *2.1 EXTENSION OF LADEVEZE DAMAGE MODEL FOR CMC MATERIALS*

This section presents the extension of (Ladeveze, P 1983)'s model for damage in CMC materials, which is given in the Appendix A, to incorporate thermal expansion and coupled thermal and mechanical effects. In the new model a linear thermal expansion coefficient is added to equation of state to account for the thermal to mechanical coupling effects. The mechanical to thermal effects are introduced by assuming an internal heat generation due to residual strain effects. The constitutive relations of the newly obtained model consists of two parts, a part related to the thermal effects and another part referring to temperature dependent physics based constitutive relations for cracking in CMCs.

#### 2.1.1 THERMAL EFFECTS

The evolution of the temperature during thermo-mechanical deformation is assumed to be governed by a linear heat equation using the assumption of transverse heat transfer:

$$\rho C \frac{d\theta}{dt} - k \Delta \theta = Q \qquad (1)$$

where $\rho$ is the density, $C$ is the specific heat capacity, $k$ is the heat conductivity, $\Delta$ denotes the Laplacian operator symbol, and $Q$ is the energy dissipated per unit time and volume, which is defined by

$$Q = \omega\, \boldsymbol{\sigma} : \boldsymbol{D}^r = \boldsymbol{\sigma} : (\boldsymbol{D}^{r_1} + \boldsymbol{D}^{r_2}) \qquad (2)$$

where $\boldsymbol{\sigma}$ is the Cauchy stress, $\boldsymbol{D}^r$ is the residual strain tensor which includes both the effects of residual strain associated to each of the direction of the tows of the woven composite, and $\boldsymbol{\omega}$ is the fraction of the dissipation energy that is converted into heat.

Note that for problems involving high strain rates, a non-conducting (adiabatic) temperature change can be assumed, as suggested by (Bammann 1993). Modeling thermal effects in high strain rates' problems also include the assumption that an important part (for instance 90%) of the inelastic work is dissipated as heat. This simple solution can permit non-isothermal solution by a FE code that is not fully coupled with the energy balance equation.

### 2.1.2 CMC DAMAGE MODEL

The constitutive equations of the material model are written in the context of linearized elasticity, in Lagrangian formulation. They consist of temperature dependent expressions for the elastic and residual deformations ("yield criterion" and associated flow rule) as well as the evolution equation of the internal parameters (matrix and fiber damage evolution laws and hardening parameters' evolution laws).

The starting point of the theoretical equations of the extended CMC material model is the classical additive decomposition of the total strain:

$$\boldsymbol{D} = \boldsymbol{D}^e + \boldsymbol{D}^r + \boldsymbol{D}^{th} \qquad (3)$$

where $\boldsymbol{D}^{th}$ is the thermal strain, $\boldsymbol{D}^e$ and $\boldsymbol{D}^r$ are the elastic and residual strains. Each of the term in this decomposition is defined as below.

#### 2.1.2.1 THERMAL STRAIN PART

The thermal strain part is defined by the classical relationship

$$\boldsymbol{D}^{th} = \alpha(\theta - \theta_{ref})\boldsymbol{I} \qquad (4)$$

where $\alpha$, $\theta$, $\theta_{ref}$, $\boldsymbol{I}$ are the scalar thermal expansion coefficient, the current temperature, the reference temperature, and the second order identity tensor, respectively.

#### 2.1.2.2 ELASTIC STRAIN PART

The elastic strain term in Eq. (3) is given by a nonlinear expression relating the strain with the Cauchy stress. We assume that the elastic moduli, $\mathbb{A}_0$, and $\mathbb{B}$ are independent of the temperature, which is a rather crude assumption since the temperature rise can degrade the material stiffness (this is only true, in the case of CMCs, when the temperature is very high). The elastic strain part is given as:

$$\boldsymbol{D}^e = \mathbb{A}:\boldsymbol{\sigma}^+ + \mathbb{A}_0:\boldsymbol{\sigma}^- + \mathbb{B}:\boldsymbol{\sigma} \qquad (5)$$

The similarity between of Eq. (A.3) of the Appendix A and Eq. (5) is only apparent since the Cauchy stress in Eq. (5) implicitly dependents upon the temperature.

### 2.1.2.3 RESIDUAL STRAIN PART

The residual strain is described using two uncoupled classical formulations of "associated plasticity" with isotropic hardening (one for each tow direction of the composite: the longitudinal and the transversal directions):

$$\boldsymbol{D}^r = \boldsymbol{D}^{r_1} + \boldsymbol{D}^{r_2} \qquad (6)$$

#### 2.1.2.3.1 RESIDUAL STRAIN TERM IN THE LONGITUDINAL DIRECTION

In the longitudinal tow, the following effective and equivalent stresses are assumed (as defined in (Ladeveze, P 1983)'s original model)

$$\begin{cases} \overline{\boldsymbol{\sigma}_1} = \boldsymbol{P}_1\, \mathbb{A}\mathbb{A}_0^{-1} \boldsymbol{\sigma}^+ \\ \overline{\sigma}_1^{eq} = \sqrt{tr(\overline{\boldsymbol{\sigma}}_1^2)} \end{cases} \qquad (7)$$

with $\boldsymbol{P}_1 = \begin{matrix} 1 & 0 & 0 \\ 0 & 0 & 0 \\ 0 & 0 & \beta \end{matrix}$

where $\beta$ is the material model parameter defining the influence of the shear on the inelastic deformation. The temperature-depend yield surface is defined as

$$f = \sigma_1^{eq} - R(r_1, \theta) - R_0(\theta) \qquad (8)$$

where $r_1$ is the cumulative residual strain and $R$ is the temperature-dependent hardening function which must be calibrated experimentally, the latter is defined by:

$$R(\theta, p) = K(\theta) p^{1/m(\theta)} \tag{9}$$

with

$$\begin{cases} K(\theta) = C_1(\theta - \theta_{ref}) + C_2 \\ m(\theta) = C_3(\theta - \theta_{ref}) + C_4 \\ R_0(\theta) = C_5(\theta - \theta_{ref}) + C_6 \end{cases} \tag{10}$$

Note that Arrhenius temperature dependence-type functions could have been adopted in (Eq. 10). However, for the CMC material of interest to this work, experimental observations have demonstrated that the hardening regime does not drastically increase with temperature rise (at least up to 1200°C), which may otherwise have justified the choice of Arrhenius type of hardening dependent functions. Based on this observation, the proposed temperature dependence functions were chosen.

The inelastic flow rule defines the following evolution law:

$$\begin{cases} \dot{\overline{D}}^{r_1} = \dot{\mu} \dfrac{\partial f}{\partial \sigma} = \dot{\mu} \dfrac{\overline{\sigma}_1}{\sqrt{tr(\overline{\sigma}_1^2)}} \\ \dot{r}_1 = -\dot{\mu} \dfrac{\partial f}{\partial R} = \dot{\mu} \end{cases} \tag{11}$$

where $\dot{\overline{D}}^{r_1}$ is the effective residual strain rate and $\dot{\mu}$ is the residual multiplier rate, which can be calculated through the consistency condition. Finally, to keep the dissipation constant, the actual residual strain rate is given by:

$$\dot{D}^{r_1} = P_1 \, \mathbb{A} \, \mathbb{A}_0^{-1} \dot{\overline{D}}^{r_1} \tag{12}$$

2.1.2.3.2 RESIDUAL STRAIN TERM IN THE TRANSVERSE DIRECTION

The formulation for the transverse tow (of the CMC material) is similar as in the longitudinal direction. The effective and equivalent stresses are defined as below:

$$\begin{cases} \overline{\boldsymbol{\sigma}_2} = \boldsymbol{P}_2\, \mathbb{A}\mathbb{A}_0^{-1}\boldsymbol{\sigma}^+ \\ \overline{\sigma}_2^{eq} = \sqrt{tr(\overline{\boldsymbol{\sigma}}_2^2)} \end{cases} \qquad (13)$$

with $\boldsymbol{P}_2 = \begin{matrix} 0 & 1 & 0 \\ 0 & 0 & 0 \\ 0 & 0 & \beta \end{matrix}$

Here also $\beta$ represent the influence of shear on the inelastic deformation. The temperature-depend yield surface in this case is also defined by:

$$g = \sigma_2^{eq} - R(r_2, \theta) - R_0(\theta) \qquad (14)$$

where $r_2$, cumulative residual strain, and $R$, the temperature-dependent hardening function are given as in the case of the definition of the residual strain the longitudinal direction.

The associated plasticity flow rule yields the following evolution laws:

$$\begin{cases} \dot{\overline{\boldsymbol{D}}}^{r_2} = \dot{\mu}\dfrac{\partial g}{\partial \boldsymbol{\sigma}} = \dot{\mu}\dfrac{\overline{\boldsymbol{\sigma}}_2}{\sqrt{tr(\overline{\boldsymbol{\sigma}}_2^2)}} \\ \dot{r}_2 = -\dot{\mu}\dfrac{\partial g}{\partial R} = \dot{\mu} \end{cases} \qquad (15)$$

where $\dot{\overline{\boldsymbol{D}}}^{r_2}$ is the effective residual strain rate and $\dot{\mu}$ is the residual multiplier rate, which can be calculated through some consistency condition. Finally, to keep the dissipation constant (here also), the actual residual strain rate is given by:

$$\dot{\boldsymbol{D}}^{r_2} = \boldsymbol{P}_2 \mathbb{A}\, \mathbb{A}_0^{-1} \dot{\overline{\boldsymbol{D}}}^{r_2} \qquad (16)$$

The remaining elements of the constitutive relations of the CMC material model consist of (i) temperature-dependent relations for the Young and shear moduli which are used to define the stiffness of the material (ii) temperature dependent damage hardening coefficient as in the original model of Ladeveze. The equations for the evolution of the matrix and fibers damage law remain formally the same as in the original CMC material model.

### 2.1.2.4 TEMPERATURE DEPENDENT YOUNG AND SHEAR MODULUS

The temperature dependent orthotropic Young and Shear moduli are defined as:

$$E_i = E_i \left(1 - E_{temp}(\theta - \theta_{ref})\right) \tag{17}$$

$$G_i = G_i \left(1 - E_{temp}(\theta - \theta_{ref})\right) \tag{18}$$

for $i = \{1,2,3\}$ and $E_i$ and $G_i$ representing the Young modulus and the shear modulus of the composite in the principal direction of the composite.

### 2.1.2.5 TEMPERATURE DAMAGE HARDENING COEFFICIENT

The temperature-dependent matrix and fiber degradation yield stress limits are given as:

$$\begin{cases} \sigma_{m0} = \sigma_{m0}(1 - C_{\theta m}(\theta - \theta_{ref})) \\ \sigma_{m1} = \sigma_{m1}(1 - C_{\theta m}(\theta - \theta_{ref})) \\ \sigma_{f0} = \sigma_{f0}\left(1 - C_{\theta f}(\theta - \theta_{ref})\right) \\ \sigma_{f1} = \sigma_{f1}\left(1 - C_{\theta f}(\theta - \theta_{ref})\right) \\ \sigma_{f2} = \sigma_{f2}(1 - C_{\theta f}(\theta - \theta_{ref})) \end{cases} \tag{19}$$

## 3.0 INTEGRATION OF THE CMC MODEL INTO ABAQUS FE CODE

### 3.1 THERMAL PART OF THE MODEL

The thermo-mechanical constitutive relations of the CMC material model presented in Section 2 were implemented into the implicit version of the ABAQUS FE code. A fully coupled thermal-displacement simulation was used to analyze and solve both the thermal and mechanical response of the CMC material. The deformation of the CMC material involves heating to inelastic deformation. Thus, the thermal and mechanical deformation must be obtained at the same time. In this formulation, the stress depends on the temperature distribution and the temperature distribution depends on the stress solution, as each of them is related to the other one and must be obtained simultaneously. In the thermal-stress analysis in ABAQUS/STANDARD the temperatures are integrated using a backward difference scheme, and the non-linear coupled system is solved using the Newton's method.

The exact implementation of Newton's method involves non-symmetric Jacobian matrix as provided in Eq. (20) illustrating the matrix representation of coupled equations:

$$\begin{bmatrix} \mathbb{K}_{uu} & \mathbb{K}_{u\theta} \\ \mathbb{K}_{\theta u} & \mathbb{K}_{\theta\theta} \end{bmatrix} \begin{pmatrix} \Delta \boldsymbol{u} \\ \Delta \theta \end{pmatrix} = \begin{pmatrix} \mathbb{R}_u \\ \mathbb{R}_\theta \end{pmatrix} \qquad (20)$$

where $\Delta \boldsymbol{u}$ and $\Delta \theta$ are the respective corrections to the incremental displacement and temperature, $\mathbb{K}_{ij}$ are sub-matrices of the fully coupled Jacobian matrix, and $\mathbb{R}_u$ and $\mathbb{R}_\theta$ are the mechanical and thermal residual vectors, respectively.

The solution of the system of equations Eq. (20) requires the use of a non-symmetric matrix storage and solution scheme. Furthermore, the mechanical and thermal equations must be solved simultaneously.

The total strain, the temperature, the increments of the total strain and temperature are computed from the global iterations between the time $t$ and $t + \Delta t$ and passed to the user-defined material algorithm. The temperature and its increment are used to update the temperature dependent parameters in Eq. (9), Eq. (10), Eq. (17), Eq. (18), and Eq. (19). Once this update is completed, the thermal strain part $\boldsymbol{D}^{th}$ is calculated according to Eq. (4).

The calculation of the residual strain $\boldsymbol{D}^r$ is more involved, but straightforward. First, the yield functions $f$ and $g$ in Eq. (8) and Eq. (14) representing the yield criteria in both the transversal and longitudinal directions are evaluated using the values of the stresses from the previous time step. The algorithm considers two different situations, one where the stress state is still in the elastic regime and the other one where the yield criterions (in both two direction of the composite) are met.

## 3.2 ISOTHERMAL PART OF THE CMC MODEL

### 3.2.1 ELASTIC REGIME ALGORITHM

When the yield criterion is not met in the two directions (i.e., the CMC material is still in the elastic regime), the elastic strain is computed by subtracting the thermal deformation from the total deformation; the Cauchy stress and the damage state variables are calculated following the algorithm developed by (Genet, M; Marcin, L.; Ladeveze, P 2013) for modeling a family of anisotropic damage with unilateral effects theory employed to model to model damage in CMC materials.

In the nonlinear elastic relationship connecting the strain with the stress Eq. (5), the damage is driven by the stress, which varies as a function of the damage; also, the state law is not reversible. Thus, the numerical algorithm of Genet et al. consists of a local loop made of embedded fixed point iterations and Newton-Raphson iterations. The fixed-point solver is used to calculate the damage state, while the Newton iterations are used to reverse the state law (which is nonlinear even when all the state variables are fixed).

#### 3.2.1.1 FIXED POINT METHOD

The fixed point algorithm given in Figure 1 is used to calculate the damage variable, which accounts for the contributions of both the matrix cracking and fibers degradation. An illustration of this algorithm along with an Aitken relaxation to accelerate the solver is also provided in Figure 2.

$$\begin{aligned}
&Initialization: j = 0;\ \boldsymbol{\alpha}^j = \boldsymbol{\alpha}^{l-1};\ \mathbb{A}^j = \mathbb{A}^{l-1};\ \mathbb{B}^j = \mathbb{B}^{l-1}\\
&\qquad\qquad Loop:\\
&Stress: \boldsymbol{\sigma}^j/\boldsymbol{\varepsilon}^e = \mathbb{A}^j \langle \boldsymbol{\sigma}^j \rangle_+^{\mathbb{A}} + \mathbb{A}_0 \langle \boldsymbol{\sigma}^j \rangle_-^{\mathbb{A}_0} + \mathbb{B}^j \boldsymbol{\sigma}^j\\
&\qquad Residual: R^j = \boldsymbol{\alpha}(\boldsymbol{\sigma}^j) - \boldsymbol{\alpha}^j\\
&ExitTest: \frac{|R^j|}{|\boldsymbol{\alpha}^j - \boldsymbol{\alpha}^{l-1}|} < tolerance\ \to exit\\
&Damage \begin{cases} \boldsymbol{\sigma}^{j+1} = \boldsymbol{\sigma}^j + R^j \\ \mathbb{A}^{j+1} = \mathbb{A}\left(\mathbb{A}^{l-1}, \boldsymbol{\alpha}^{j+1} - \boldsymbol{\alpha}^{l-1}\right) \\ \mathbb{B}^{j+1} = \mathbb{B}\left(\mathbb{B}^{l-1}, \boldsymbol{\alpha}^{j+1} - \boldsymbol{\alpha}^{l-1}\right) \end{cases}\\
&\qquad EndLoop: j = j + 1
\end{aligned}$$

**Figure 1. Fixed Point Method**

with

$$\sigma^{j+1} = \sigma^j + s^j R^j$$

$$with\, s^j = \begin{cases} 1 \text{ if } j = 0 \\ -s^{j-1} \dfrac{R^{j-1}}{R^j - R^{j-1}} \text{ if } j > 0 \end{cases}$$

**Figure 2. Aitken Convergence Acceleration Factor**

### *3.2.1.2 NEWTON-RAPHSON METHOD*

The solution for the nonlinear equation of state is solved using the Newton-Raphson method also with an Aitken relaxation operator to accelerate the convergence of the solver. The methodology is illustrated in Figure 3 and Figure 4.

$$Initialization: k = 0;\ \sigma^k = \sigma^{l-1}$$
$$Loop:$$
$$Residual:$$
$$R^k = \varepsilon^e - \mathbb{A}\langle\sigma^k\rangle_+^{\mathbb{A}} - \mathbb{A}_0\langle\sigma^k\rangle_-^{\mathbb{A}_0} + \mathbb{B}\,\sigma^k$$
$$Exit\, test\, \|R^k\| < tolerance \rightarrow Exit$$
$$Stress: \sigma^{k+1} = \sigma^k + D^k R^k$$
$$EndLoop$$

**Figure 3. Newton-Raphson Method**

with

$$D^k = \begin{cases} \left(\mathbb{A}^{l-1} + \mathbb{B}^{l-1}\right)^{-1} if\, tr(\sigma^k) > 0 \\ \left(\mathbb{A}_0 + \mathbb{B}^{l-1}\right)^{-1} if\, tr(\sigma^k) < 0 \end{cases}$$

**Figure 4. Convergence Acceleration Factor**

### 3.2.2 INELASTIC REGIME ALGORITHM

When the yield criteria in both the transversal and longitudinal direction are met, a radial return algorithm is used to calculate the "plastic" multiplier for each direction of the material. Then the residual strain in each of the two cases is evaluated to obtain to total residual strain. The latter is added to the thermal part of the strain and the result is subtracted to the total deformation to obtain the elastic strain part.

After the elastic strain is computed the nonlinear equation of state Eq.(5) is solved for the damage stiffness matrices and the Cauchy stress based on the fixed point and the Newton-Raphson methods as already described in previous sections.

## 4.0 DETERMINATION OF THE CMC MATERIAL MODEL CONSTANTS

The CMC material model requires 47 input parameters which relate to the physical, mechanical, and thermal properties of the material. A strategy based on the proposals of (Gasser, Ladeveze and Poss 1996) and (Letombes, S. 2005) to identify each of these parameters is used; in this the parameters are grouped into the following categories:

1. The matrix degradation parameters,
2. The fiber degradation parameters,
3. The residual strain parameters,
4. The damage hardening parameters related to the matrix and fiber damage laws,
5. The temperature dependent residual strain parameters,
6. The temperature-dependent damage hardening parameters related to matrix and fiber damage laws, and
7. The temperature-dependent Young and shear moduli coefficient.

Table 1 describes each of the model parameters.

**Table 1. CMC Model Parameters**

| Material Parameter | Category | Description |
|---|---|---|
| $a$ | 1 | Combines the isotropic and directional damages |
| $n$ | 1 | Drives the matrix tensile damage anisotropic effects |
| $b$ | 1 | Drives the matrix shear damage anisotropic effects |
| $c$ | 2 | Define the weighting factor for the fiber damage |
| $c'$ | 2 | Defines the fibers tensile damage anisotropic effects |
| $e$ | 2 | Defines the onset of the fibers damage |
| $e'$ | 2 | Defines the fibers shear damage anisotropic effects |
| $C_1$ | 3 | Controls the hardening of the composite at RT |
| $C_2$ | 3 | Controls the hardening of the composite at ET |
| $C_3$ | 3 | Controls the hardening slope of the material at RT |
| $C_4$ | 3 | Controls the hardening slope of the Composite at ET |

| Material Parameter | Category | Description |
|---|---|---|
| $C_5$ | 3 | Temperature dependent yield limit |
| $C_6$ | 3 | Yield limit of the composite for ambient temperature |
| $\beta$ | 3 | Defines the influence of shear on the inelastic response of the composite |
| $\sigma_{\alpha m0}$ | 4 | Defines the stress at which matrix degradation hardening starts |
| $\sigma_{\alpha m1}$ | 4 | Defines the stress at which matrix degradation hardening saturates |
| $a_{\alpha m1}$ | 4 | Defines the matrix degradation hardening coefficient |
| $\sigma_{\alpha f0}$ | 4 | Defines the fiber strength before any fiber degradation starts |
| $\sigma_{\alpha f1}$ | 4 | Defines the fiber strength at which the fiber degradation starts |
| $\sigma_{\alpha f2}$ | 4 | Defines the fiber strength after which the fiber fails |
| $a_{\alpha f1}$ | 4 | Defines the fiber degradation hardening coefficient |
| $a_{\alpha f2}$ | 4 | Defines the fiber degradation hardening coefficient |
| $E_1$ | NA | Anisotropic Young Modulus direction 1 |
| $E_2$ | NA | Anisotropic Young Modulus direction 2 |
| $E_3$ | NA | Anisotropic Young Modulus direction 3 |
| $\nu_{12}$ | NA | Anisotropic Poisson Ratio direction 1 |
| $\nu_{13}$ | NA | Anisotropic Poisson Ratio direction 2 |
| $\nu_{23}$ | NA | Anisotropic Poisson direction 3 |
| $G_{12}$ | NA | Anisotropic Shear Modulus in the x-y plane |
| $G_{13}$ | NA | Anisotropic Shear Modulus in the x-z plane |
| $G_{23}$ | NA | Anisotropic Shear Modulus in the y-z plane |
| $\theta_{ini}$ | NA | Initial temperature |
| $\theta_{ref}$ | NA | Reference temperature |
| $\alpha$ | NA | Thermal expansion coefficient |

| Material Parameter | Category | Description |
|---|---|---|
| $\omega$ | NA | Inelastic heat fraction |
| $C_v$ | NA | Heat capacity |
| $C_{\theta m}$ | 6 | Temperature-dependent matrix degradation yield stress constant |
| $C_{\theta f}$ | 6 | Constant in the temperature dependent fiber yield stresses |
| $E_{temp}$ | 7 | Young and shear moduli temperature-dependent constant |
| $\rho$ | NA | Density of the composite (use only for thermomechanical analysis) |

### 4.1.1 PARAMETERS RELATED TO MATRIX DEGRADATION

The matrix cracking evolution law contains three coefficients $(a, b, n)$ which are associated with the degradation of the matrix material. These parameters can be determined from data obtained using multi-axial and non-proportional tests. In the absence of these tests, a simple interpretation suggested by (Letombes, S. 2005) may be enough; this interpretation is presented below.

The evolution equation for matrix cracking corresponds to a combination of two types of damage, isotropic and directional, which can be associated with an independent and oriented micro-cracking of the matrix, respectively. The coefficient $a$ is a weighting parameter reflecting the anisotropic character of the matrix cracking evolution law, where $a =1$ represent an isotropic damage-type tensor. For $a =1$, it is possible to show that the matrix cracking evolution law is independent of the damage measure direction for any arbitrary direction and intensity of the load. With $a=0$ and with $n$ set to a high value, a directional behavior is obtained, where the value of $n$ accounts for direction effects.

### 4.1.2 PARAMETERS RELATED TO FIBER DEGRADATION

Parameters related to the fiber degradation $(e, e')$ can be identified using multi-axial non-proportional data and analytical expressions for some a priori known components of the fiber damage compliance tensor $\mathbb{A}_{f_1}$ (in the direction 1 of the fibers), for instance in quasi-monotonic tension-compression cyclic or tension-tension cyclic tests, at RT, at 0°C and 45°C with respect to the fiber direction. As for the parameters $(c, c')$, which are also related to fiber cracking, their identification is based, here also, on some crude arguments that are related to fibers' cracking appearance.

Several works pertaining to different SAFRAN Group CMC materials including those of (Forio, P 2000) have demonstrated that during the loading of the CMC in the fibers direction, longitudinal fibers degrade after cracks in transversal yarns (i.e., group of fibers) and in the intra-yarns matrix saturate. Because of the lack of information on the time of the initiation of damage mechanisms and the damage intensity, it is assumed that the damage in the transversal fibers is calculated when the mechanical load is oriented by longitudinal fibers. With this assumption, intra-yarn matrix cracking appears first preventing cracks that are parallel to fibers in the transversal yarns to develop; this choice enforces $c$=0.

The parameter $c'$ mainly defines the ratio between the damage originating from longitudinal and transversal yarns during tests that are performed in the fibers direction. If the value of $c'$ is assumed to be small, it reflects the influence of fiber-matrix interface de-cohesion.

Another choice for $c'$, which involves the time of the appearance of different cracks network, is possible. This other choice does not modify the macroscopic damage, but the cracking distribution. The choice between these two types of identification requires information such as microscopic information on the density of the cracking of the different elements of the composite.

### 4.1.3 Residual Strain Parameters

For isothermal analysis, the residual strain constitutive relations involve four different material constants, $\beta$, $R_0$, $K_y$, $m_y$ that are described in Table 1. The parameter $\beta$, which represents the influence of the shear damage on inelasticity, is set to some default value, while $R_0$ is the yield limit for the composite that can be obtained from a monotonic tension test. The two other parameters are to be determined from macroscopic tensile-tensile cycling with a few load/unload cycles to measure the residual deformations. The identification of these parameters follows an iterative approach in several steps. First, K&C defines from the load/unload curve the residual strains. Then, these residual deformations are modeled using a stress dependent function. These approximations of the residual strains allow extracting the elastic deformations from the total deformation and thus, obtaining the damage during the loading history. The estimated stress-strain curve is used to fit the constants $K_y$, $m_y$.

Figure 5a shows stress-strain plots generated with CMC material model with parameters that turn-off the residual strain effects. Figure 5b shows the effect of residual strain, see bottom left corner of the figure.

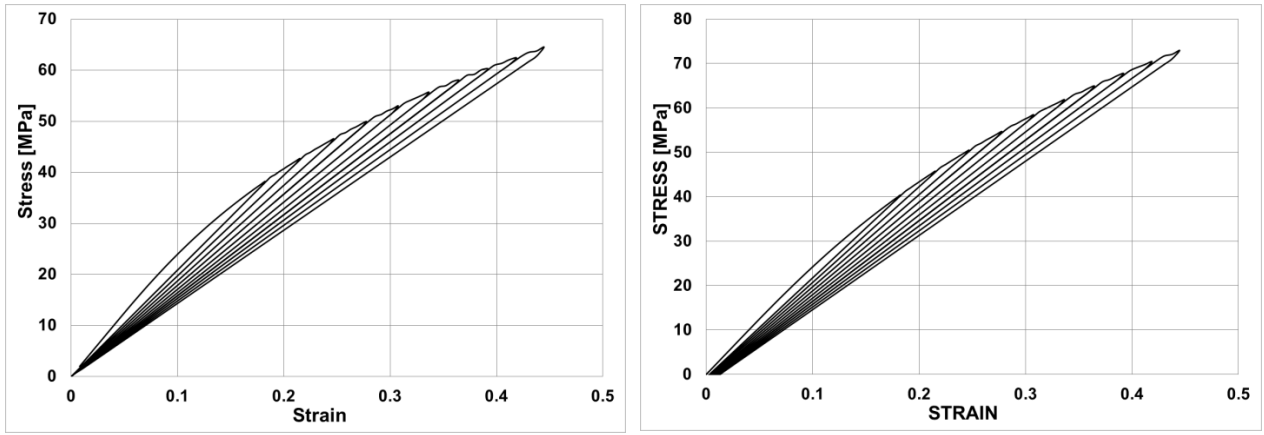

(a) With no residual strain          (b) With residual strain

**Figure 5. Effect of Residual Strain Parameters on Material Response**

### 4.1.4 TEMPERATURE-DEPENDENT YOUNG AND SHEAR MODULI

A temperature-dependent parameter for orthotropic Young and Shear moduli, $E_{temp}$, is introduced and fitted to the experimental curve at high temperature. Figure 6 illustrates the effect of this feature, which shows the impact of temperature on the mechanical properties.

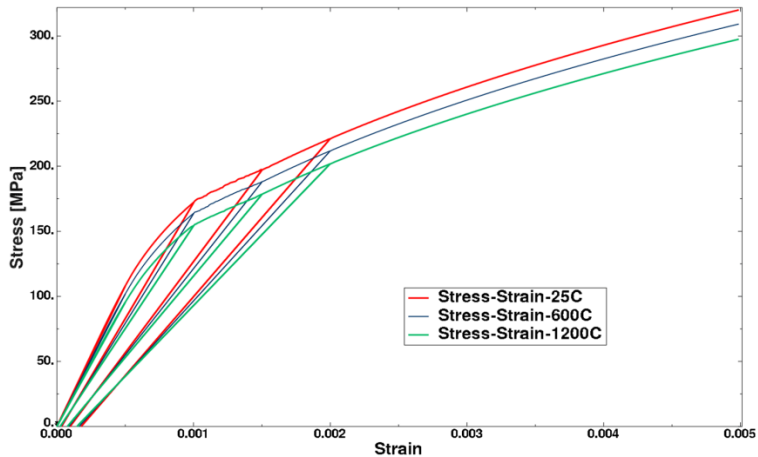

**Figure 6. Temperature Dependence of Mechanical Properties**

### 4.1.5 DAMAGE HARDENING PARAMETERS FOR MATRIX AND FIBER DAMAGE

Identifying the parameters for damage hardening can be quite involved: a piecewise identification for increasing values of the matrix and fibers damage variables is necessary. It requires using tension-tension or tension-compression cyclic data in the direction of the fibers. Experimental data with a few load/unload cycles which represent sufficiently the behavior of the composite will be considered. Residual strains will be determined from each of the cycles of the experiment; then these strains will be fitted with a stress dependent function. This approximation will be helpful to extract residual deformations from the total deformation; as a result, the damage behavior of the composite can be obtained for the entire loading process. Subsequently, the elastic deformation may be obtained by subtracting the estimated inelastic deformations from the total experimental deformations. To obtain a fit for matrix and fiber damage parameters, an analytic expression for the compliance tensor component in the direction of the longitudinal fiber that depends on the damage hardening parameters will be defined.

The initial yield limit of the composite is close to the stress at which the longitudinal fiber is damaged. Its value can be obtained from a monotonic stress-strain data for the composite.

Figure 7 shows the effect of these hardening parameters. The plot shows a stress-strain response for CMC with and without fiber degradation.

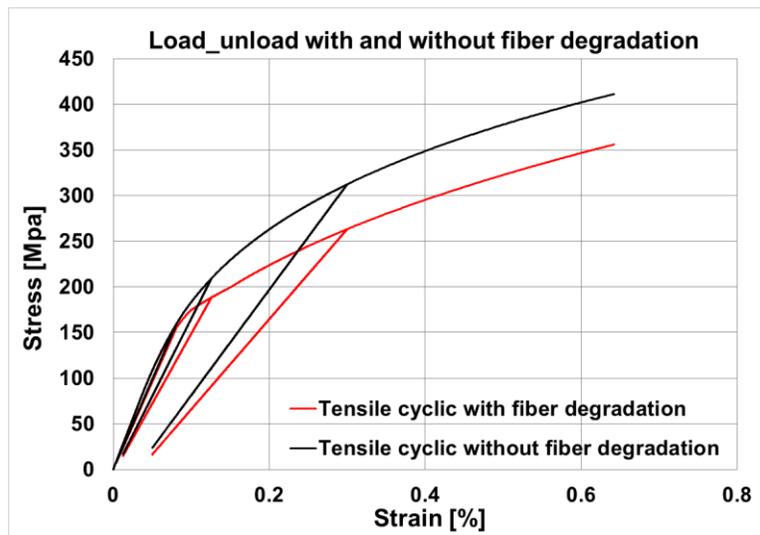

**Figure 7. Effect of Fiber Degradation on the Material Response**

### 4.1.6 TEMPERATURE-DEPENDENT DAMAGE HARDENING PARAMETERS RELATED TO THE MATRIX AND FIBER DAMAGE LAWS

Two parameters for temperature dependency are related to the laws of matrix and fiber degradation; these parameters are defined as $C_{\theta m}$ and $C_{\theta f}$. For now, due to the lack of experimental data at high temperatures, these constants are set to some default constants.

### 4.1.7 TEMPERATURE-DEPENDENT RESIDUAL STRAIN PARAMETERS

The residual strain effects constitutive relations at ET are defined as linear functions of the temperature; these relations include six temperature-dependent constitutive constants: $C_1$, $C_2$, $C_3$, $C_4$, $C_5$, $C_6$ which are described in Table 1. The identification of these constants is based on a crude approach due to the lack of experimental data. The approach is based on a trial-and-error method and consists of comparing the results predicted by the CMC damage model for a single finite element loaded in tension cyclic (with few cycles) with the SAFRAN Group CMC material true stress vs. true strain experimental curve at ET (1200°C). A summary of the obtained constants using this methodology is provided in Table 2.

Note that, with respect to the isothermal CMC damage model, two additional ($C_6$, $C_5$) constants were fitted. Figure 6 demonstrates the temperature effects on the response of a laboratory scale specimen loaded in tension cycling.

### 4.1.8 CALIBRATED PARAMETERS FOR SAFRAN GROUP CMC MATERIAL

For the SAFRAN Group material, the calibrated CMC material model parameters used in the verification and validation calculations are provided in Table 2. Some of the parameters were calibrated with test data, if the appropriate test data was available. The input parameters consider elastic modulus, shear modulus, and Poisson ratio in three different directions to account for orthotropic material behavior. The thermal conductivity, coefficient of thermal expansion, and the thermal effect on the elastic modulus as well as the strength of the matrix and fiber materials are considered with the "Conduct, $\alpha$, $E_{temp}$, $C_{\theta m}$, and $C_{\theta f}$" parameters.

**Table 2. Material Model Parameters for SAFRAN Group CMC Material**

| Parameter | Value | Parameter | Value | Parameter | Value |
|---|---|---|---|---|---|
| $\rho$ | 1.5e-09 | $E_{temp}$ | 1.e-4 | $a_{\alpha f1}$ | 1.67e-06 |
| $k$ | 1.e-03 | $\beta$ | 2. | $a_{\alpha f2}$ | 6.7821e-06 |
| $\theta_{ini}$ | 25. | $C_1$ | 0. | $a$ | 0.1 |
| $\theta_{ref}$ | 25. | $C_2$ | 380.e3 | $n$ | 2. |
| $\alpha$ | 1e-06 | $C_3$ | 0. | $b$ | 2 |
| $C_v$ | 400.e+06 | $C_4$ | 1.25 | $c$ | 0. |
| $\omega$ | 0.9 | $C_5$ | -0.068 | $c'$ | 3. |
| $E_1$ | 221.e3 | $C_6$ | 200. | $e$ | 0.1 |
| $E_2$ | 221.e3 | $\sigma_{\alpha m0}$ | 70. | $e'$ | 9. |
| $E_3$ | 221.e3 | $\sigma_{\alpha m1}$ | 200. | Epsfp | 1.e-08 |
| $\nu_{12}$ | 0.04 | $C_{\theta m}$ | 0.e-4 | Epsnrs | 1.e-08 |
| $\nu_{13}$ | 0.04 | $\sigma_{\alpha f0}$ | 130. | Vnewdt | 0.5 |
| $\nu_{23}$ | 0.04 | $\sigma_{\alpha f1}$ | 200. | Nplan | 0. |
| $G_{12}$ | $\dfrac{E_1}{2(1+\nu_{12})}$ | $\sigma_{\alpha f2}$ | 34000. | Nlgeom | 0. |
| $G_{13}$ | $\dfrac{E_2}{2(1+\nu_{13})}$ | $C_{\theta f}$ | 0.e-4 | M | 0 |

| Parameter | Value | Parameter | Value | Parameter | Value |
|---|---|---|---|---|---|
| $G_{23}$ | $\dfrac{E_3}{2(1+\nu_{23})}$ | $a_{\alpha m1}$ | 3.83474e-07 | | |

## 5.0 VALIDATION AND VERIFICATION OF THE CMC MATERIAL MODEL

### 5.1 MODEL CALIBRATION VERIFICATION

The model parameters were calibrated primarily using tensile and tensile load-unload test data at RT and ET. Verification calculations were performed using single solid elements as well as multi-element models of the test specimens. Figure8 shows the ABAQUS FE model of the specimen geometry used in the verification calculations. The geometry is representative of the geometry used in the lab tests. An example comparison of the results for the simple tension and tensile load-unload calculations and tests are shown in Figure 9 and Figure 10for (room temperatures) RTs and (elevated temperatures) ETs. The results indicate an excellent agreement between the tests and the computations with the CMC material model for the single elements. For the multi-element specimen models, the results also show good comparisons but the stress is generally over-predicted. The over-prediction may be due to the lack of information available in the open literature regarding the tests conducted for the CMC materials. For example, no information is available regarding the grips used in the tests, the manner the load is applied to the specimens, or the type/location of the strain gages used. This makes it difficult for us to accurately model the boundary conditions, input loads, and to post-process the results in a consistent manner. Another reason for the over-prediction may be due to the material model integration or the way the FE model is formulated.

The stress in the stress vs. strain curve predicted by the CMC damage model for the dog-bone specimen loaded in tension was obtained by summing up the reaction forces at all the nodes located on the edge face of the specimen where the displacement load was applied and dividing the resulting total force by the cross-section area of the specimen. The strain was obtained based on a 25-mm long section located in the specimen gauge section. The strain was computed from the displacement of two symmetric nodes located on the opposite sides of the chosen section of the specimen gauge. The section length computed from the two nodes was then divided by the initial section length to obtain the strain at each time increment.

For the rectangular specimens (with and without an open-hole), the stress in the stress vs. strain curve for the open hole tests was computed by summing up the reaction forces at all the nodes located at the face of the specimen where the displacement load was applied and dividing the obtained force by the area of this face. The strain was obtained by choosing two symmetrical nodes in the longitudinal diameter (of the hole) direction. Initially, the distance between these two nodes is 40-*mm*. The displacement of each of the two nodes is stored at each time increment and used to update the distance between the two chosen nodes. The newly computed distance between the two nodes is then divided by the initial distance between the two nodes at each time increment gives the strain.

Figure 11provides a comparison between the results predicted by the CMC-DAMAGE model and the experimental data for a dog bone specimen subjected to an incremental tension and compression loads. The comparisons show a reasonable agreement between the simulation results and the data.

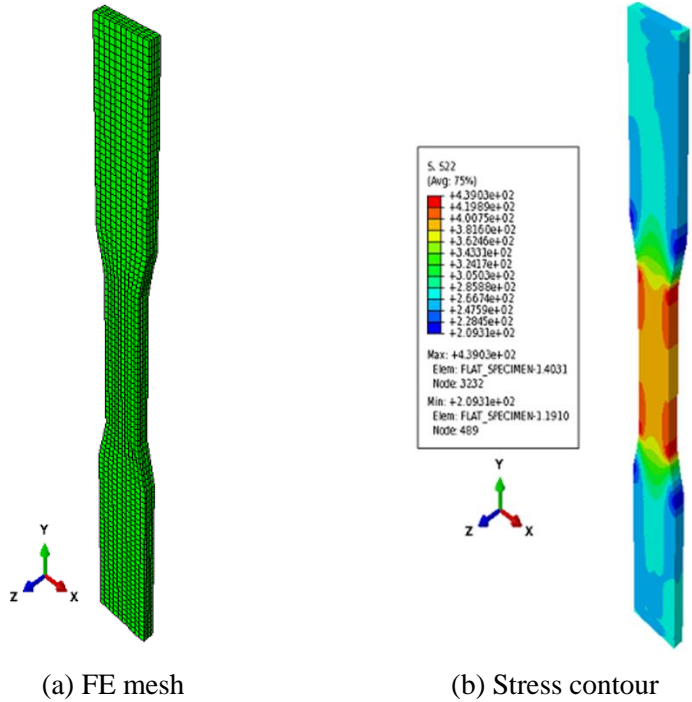

(a) FE mesh          (b) Stress contour

**Figure 8. ABAQUS FE Multi-Element Model of Test Specimen Used for Tension and Load-unload Computations**

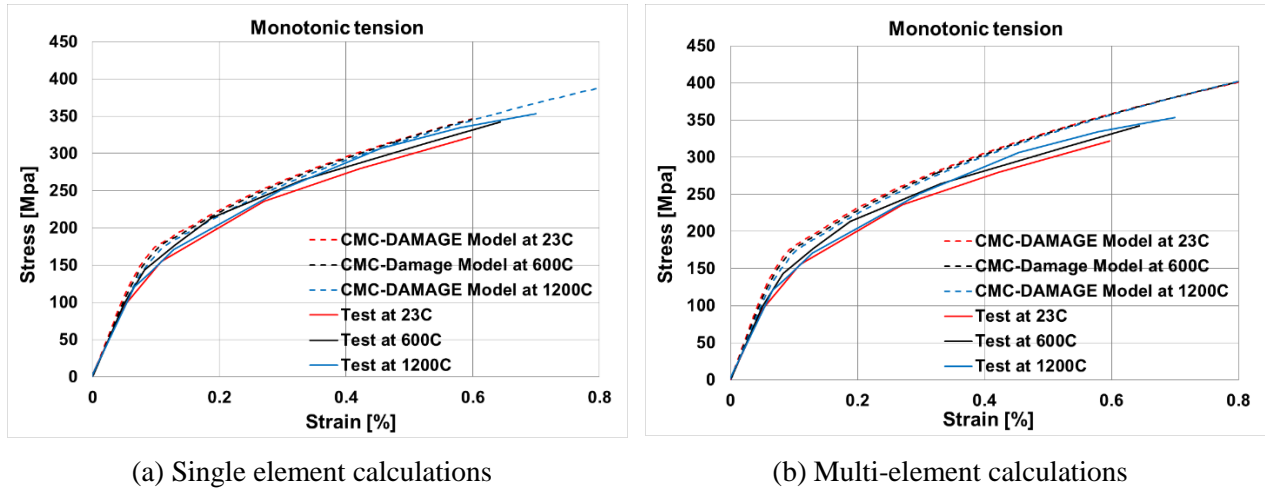

(a) Single element calculations          (b) Multi-element calculations

**Figure 9. Monotonic Tension Comparisons at RT and ETs**

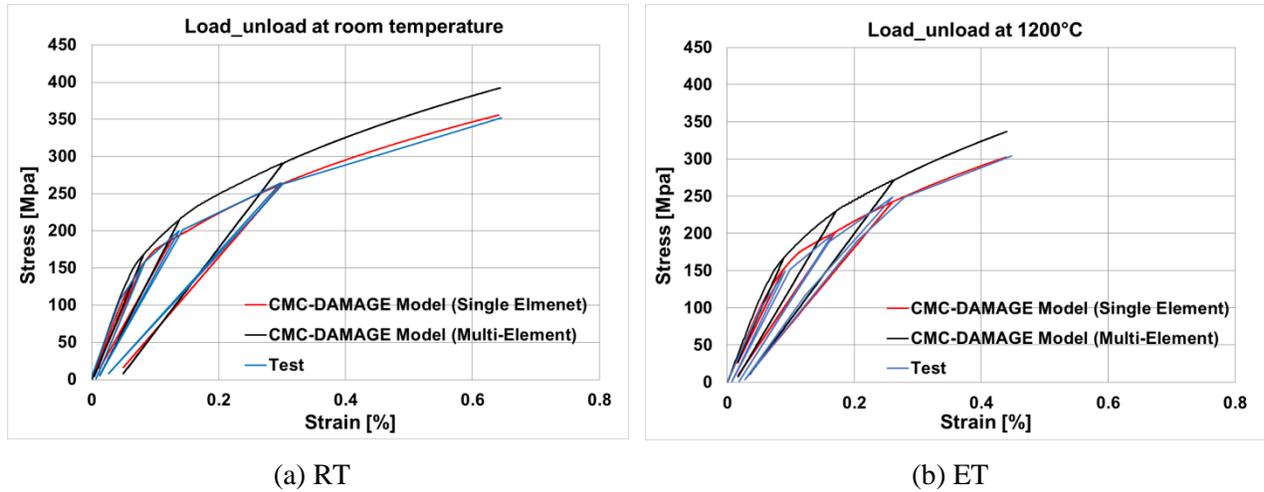

(a) RT  (b) ET

**Figure 10. Tension Load-unload Comparisons**

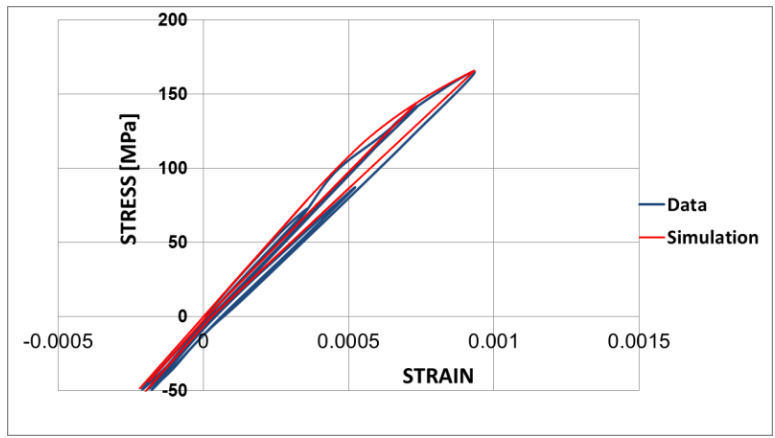

**Figure 11. Tension-compression Cyclic Comparisons**

## 5.2 MODEL VALIDATION

A limited validation of the model was performed using test data for an open-hole test. The test consisted of a rectangular-shaped specimen subjected to a monotonic tension force. Three variations of the test were conducted: 1) no hole, 2) with a 4-mm hole, and 3) with a 6-mm hole. The SAFRAN material used in these tests may not have been identical to the material used in the calibration effort but the material appears to display similar properties. The geometry and stress contours of the specimens are illustrated in Figure 12. The stress contours show the $\sigma_{22}$ stress. Predictions with the CMC model showed relatively similar results and a similar pattern as the hole size increases, see the stress-strain plots in Figure 12.

These results would likely improve (see Figure 13) if better CMC data was available, but clearly demonstrate that the model captures the nonlinearity exhibited by these tests. These results also demonstrate the clear need for a consistent and well-characterized set of CMC data by which to both calibrate the CMC material model's parameters and to validate its capability. This need is addressed in future work.

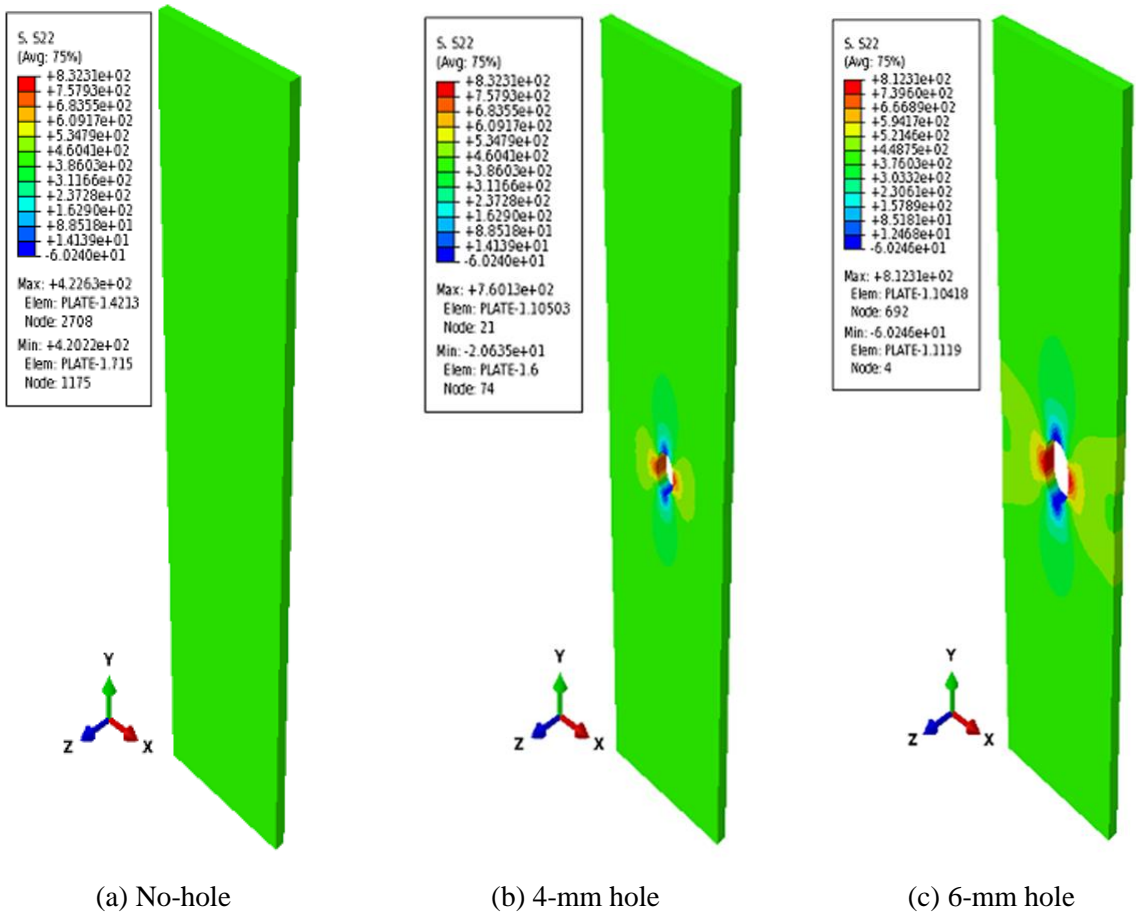

(a) No-hole  (b) 4-mm hole  (c) 6-mm hole

**Figure 12. Open-hole ABAQUS FE Predictions**

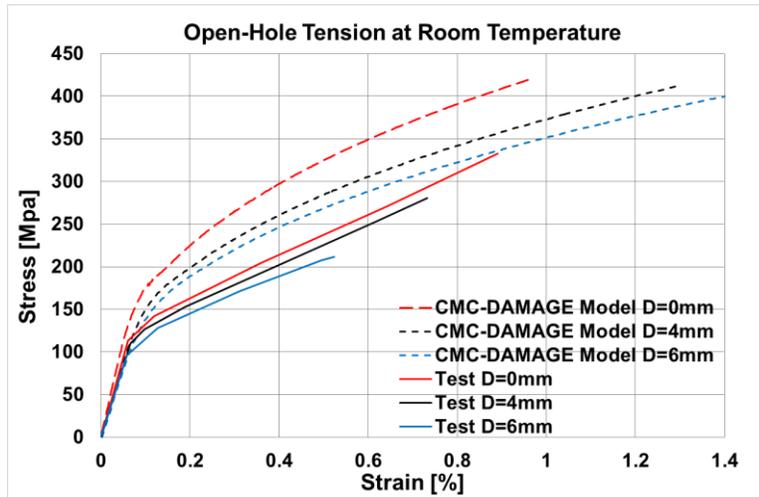

**Figure 13. Prediction of Open-Hole Tensile Tests**

**6.0 CONCLUSION**

A model for CMC materials including thermo-mechanical coupling effects that extends the one develop by Ladeveze and co-workers is presented and incorporated into ABAQUS FE code as a user material software. A procedure to obtain the CMC material model parameters for a *SiC/SiC* composite developed by SAFRAN Group and which is based on a previous work by (Letombes, S. 2005) is also provided and the material model parameters given. A verification and validation processes were conducted to check the accuracy and robustness of both the numerical implementation of the model and its ability to represent typical thermo-mechanical behaviors of CMC materials subjected to thermal and coupled thermo-mechanical loads.

## 8.0 APPENDIX A

The proposed constitutive material model for modeling CMCs in extreme mechanical, thermal, and thermomechanical environments is based on the thermodynamics of irreversible processes with internal state variables approach and accounts for inelasticity, damage and fracture behaviors. The inelastic part of the model is obtained by assuming an associated yield function obeying the normality rule, while the damage part is based on an anisotropic and unilateral damage theory developed, some years ago, by (Ladeveze, P 1983) and used in the context of the design of a new generation of high temperature CMCs. An ad hoc fracture criterion was developed for the simulation of CMC material problems involving failure.

### *8.1  INTERNAL ENERGY DENSITY FUNCTION*

The approach has the particularity of treating each damage mechanism (inter-yarn cracking, intra-yarn longitudinal cracking and intra-yarn transversal cracking) separately. The main constitutive equation of this model consists of a potential of energy density function which is expressed in terms of the stress tensor and divided into three parts: 1) a first part which is active only in traction and which takes into account the damage state, 2) a second part which is active only in compression and which is independent of the damage state, and 3) a third part which is always active and also involves some damage. Thus, this potential can be expressed as:

$$\Psi = \frac{1}{2} tr(\mathbb{A}\,\boldsymbol{\sigma}^+\boldsymbol{\sigma}^+) + \frac{1}{2} tr(\mathbb{A}_0\,\boldsymbol{\sigma}^-\boldsymbol{\sigma}^-) + \frac{1}{2} tr(\mathbb{B}\,\boldsymbol{\sigma}\boldsymbol{\sigma}) \qquad (A.1)$$

where $\mathbb{A}_0$, $\mathbb{A}$ and $\mathbb{B}$ are three fourth-order tensors representing respectively the initial compliance, the damaged compliance and a compliance operator associated with shear damage. In addition, a special decomposition of the stress tensor into a positive part and a negative part is used in order to ensure the continuity of the state law (Ladeveze, P 1983):

$$\begin{cases} \boldsymbol{\sigma}^+ = \mathbb{A}^{-1/2}:\langle \mathbb{A}^{1/2}:\boldsymbol{\sigma}\rangle_+ \\ \boldsymbol{\sigma}^- = \mathbb{A}^{-1/2}:\langle \mathbb{A}^{1/2}:\boldsymbol{\sigma}\rangle_- \end{cases} \qquad (A.2)$$

where the symbol $\langle X \rangle_{+/-}$ represents the classical positive and negative parts of the quantity $X$.

#### 8.1.1  EQUATION OF STATE

The positive part and the negative part must be taken in that special sense (and not in the classical sense) in order to avoid a discontinuous state law. Thus, the state law becomes:

$$\mathbf{D} = \frac{\partial \Psi}{\partial \boldsymbol{\sigma}} = \mathbb{A}\,\boldsymbol{\sigma}^+ + \mathbb{A}_0\boldsymbol{\sigma}^- + \mathbb{B}\,\boldsymbol{\sigma} \qquad (A.3)$$

where **D** is the total elastic strain tensor.

The energy release rates associated with the variations of internal variables $\mathbb{A}$ and $\mathbb{B}$, i.e. the thermodynamics forces, are then defined as

$$\begin{cases} \mathbb{M} = \dfrac{\partial \Psi}{\partial \mathbb{A}} = \dfrac{1}{2}\boldsymbol{\sigma}^+ \otimes \boldsymbol{\sigma}^+ \\ \mathbb{M}' = \dfrac{\partial \Psi}{\partial \mathbb{B}} = \dfrac{1}{2}\boldsymbol{\sigma} \otimes \boldsymbol{\sigma} \end{cases} \qquad (A.4)$$

An additional thermodynamic force, which is required to drive shear damage correctly, is defined as

$$\begin{cases} \mathbb{M}'' = {}^1\!/_2 \left(\boldsymbol{I}_{\pi/2}\boldsymbol{\sigma}^+\right)_{sym} \oplus \left(\boldsymbol{I}_{\pi/2}\boldsymbol{\sigma}^+\right)_{sym} \\ \boldsymbol{I}_{\pi/2} = \begin{bmatrix} 0 & -1 \\ 1 & 0 \end{bmatrix} \end{cases} \qquad (A.5)$$

### 8.1.2 DAMAGE EVOLUTION LAWS

Each degradation mechanism is associated with a damage evolution law which affects part or all of tensors $\mathbb{A}$ and $\mathbb{B}$:

$$\begin{cases} \mathbb{A} = \mathbb{A}_m + \mathbb{A}_{f_1} + \mathbb{A}_{f_2} \\ \mathbb{B} = \mathbb{B}_m + \mathbb{B}_{f_1} + \mathbb{B}_{f_2} \end{cases} \qquad (A.6)$$

*where* $\mathbb{A}_m$ and $\mathbb{B}_m$ represent inter-yarn matrix, $\mathbb{A}_{f_1}$ and $\mathbb{B}_{f_1}$ denote intra-yarn matrix cracking of the longitudinal tows, and $\mathbb{A}_{f_2}$ and $\mathbb{B}_{f_2}$ represent the inter-yarn matrix and intra-yarn matrix cracking of the transverse tows.

### *8.1.2.1 MATRIX DAMAGE LAW*

In the case of inter-yarn matrix cracking (an illustration is provided in Figure 14), the effective thermodynamics force and its maximum over time are considered, as cited below:

$$\begin{cases} z_m = [a(tr(\mathbb{M}))^{n+1} + (1-a)tr(\mathbb{M}^n)]^{1/n+1} \\ \bar{z}_m(t) = \sup_{r<t}(z_m(t)) \end{cases} \quad (A.7)$$

and the definition of the evolution equation for the damage tensor variables become

$$\begin{cases} \mathbb{A}_m = [a(tr(\mathbb{M}))^n + (1-a)\mathbb{M}^n]\dfrac{\alpha_m}{\bar{z}_m^n} \\ \mathbb{B}_m = \dfrac{\alpha_m}{z_m} b\mathbb{M}'' \end{cases} \quad (A.8)$$

where $\alpha_m$ is a function of $\bar{z}_m$ which needs to be calibrated experimentally, and *a, b,* and *n* are parameters of the model defining damage anisotropy. The other damage laws have similar expressions as shown in (Ladeveze, P 1983). The formulation for the transversal tow is analogous to the one shown in (Baranger, E.; Cluzel, C.; Ladeveze, P.; Mouret, A 2007) and (Letombes, S. 2005).

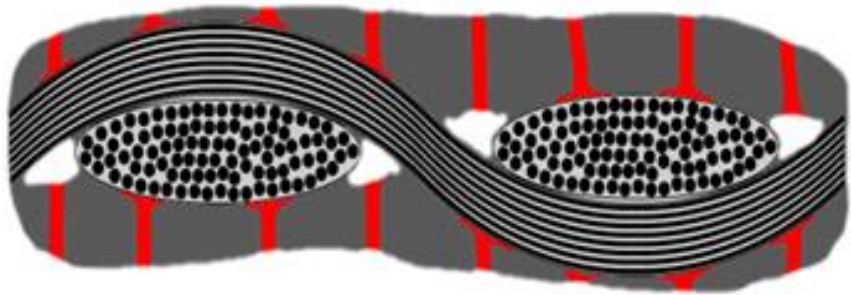

**Figure 14. Matrix Crack Network(Genet, et al. 2008)**

### 8.1.2.2 FIBER DAMAGE LAW

The damage in the yarn is due to both matrix and fiber cracking, an illustration is provided in Figure 15. The failure of the composite is related to the fiber cracking. Whether the load applied to the composite is of fatigue, oxidation, or quasi-static mechanics type, all of the modeling techniques for fracture assume that fiber cracking only appears in the ultimate time of the material service life or in the ultimate time of the applied mechanical load, (Forio, P 2000). Thus, fiber failure is activated in the modeling technique by a user-defined brittle threshold.

Similar to the modeling of matrix cracking, a damage force with the effective scalar $\bar{z}_{f_1}$ for the yarn in the direction 1 and $\bar{z}_{f_2}$ for the yarn in the direction 2 is defined as

$$z_{f_1} = \left[ \left( tr[M_1^{yarn1} \cdot \mathbb{M}] + c\, tr[M_2^{yarn1} \cdot \mathbb{M}] \right)^2 + e\left( tr[M_{12}^{yarn1} \cdot \mathbb{M}] \right)^2 \right]^{1/2} \quad (A.9)$$

with the projectors expressed in the global reference the axis of which are fixed by the yarns (1: longitudinal yarn; 2: transversal yarn):

$$M_1^{yarn1} = M_2^{yarn2} = \begin{matrix} 1 & 0 & 0 \\ 0 & 0 & 0 \\ 0 & 0 & 0 \end{matrix}, \quad M_2^{yarn1} = M_1^{yarn2} = \begin{matrix} 0 & 0 & 0 \\ 0 & 1 & 0 \\ 0 & 0 & 0 \end{matrix},$$

$$M_{12}^{yarn1} = M_{12}^{yarn2} = \begin{matrix} 0 & 0 & 0 \\ 0 & 0 & 0 \\ 0 & 0 & 1 \end{matrix} \quad (A.10)$$

For the direction of the fiber 1, the matrix $M_1^{yarn1}$ allows to extract the part of the damage force, $\mathbb{M}$ which corresponds to the cracks that are orthogonal to the direction of the fibers. Similarly, the matrix $M_2^{yarn1}$ is related to the damage corresponding to cracks that are parallel to the direction of the fibers. The matrix $M_{12}^{yarn1}$ defines the influence of the shear upon the damage.

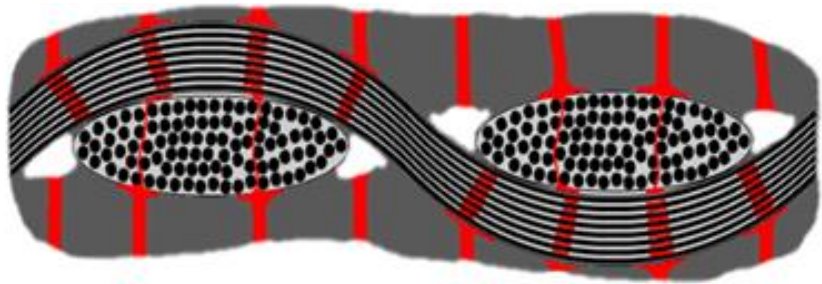

**Figure 15. Combined Matrix and Fiber Crack Networks (Genet, et al. 2008)**

The coefficients $c$ and $e$ allow to weight the influence of the projections of the forces in the direction of the fibers, and in the orthogonal direction of the fibers and the shearing on the fibers degradation.

The history in the time interval $[t_0, t]$ is accounted for by the function defined by

$$\bar{z}_{f_1}(t) = sup_{r<t}\left(z_{f_1}(t)\right) \qquad (A.11)$$

From there, the damage strain rate related to the fibers is defined as:

$$\begin{cases} \mathbb{A}_{f_1} = \dfrac{\alpha_f\left(\sqrt{\bar{z}_{f_1}}\right)}{\bar{z}_{f_1}}\left(tr\left[M_1^{yarn1}.\mathbb{M}\right] + ctr\left[M_2^{yarn1}.\mathbb{M}\right]\right)\left(M_1^{yarn1} + c'\, M_2^{yarn1}\right) \\ \\ \mathbb{B}_{f_1} = \dfrac{\alpha_f\left(\sqrt{\bar{z}_{f_1}}\right)}{\bar{z}_{f_1}} e'tr\left(M_{12}^{yarn1}\mathbb{M}'\right)M_{12}^{yarn1} \end{cases} \qquad (A.12)$$

The coefficients $c'$ and $e'$ represent the weighting factors for the anisotropy of the fiber damage in direction 1 of the fibers.

For the fibers in the second direction, similar evolution laws are utilized, using in the previous evolution equations the following projectors: $M_1^{yarn2}$, $M_2^{yarn2}$, $M_{12}^{yarn2}$

### 8.1.2.3   *DAMAGE HARDENING EQUATIONS*

Both the matrix and fiber damage laws involve damage strain variables $\alpha_f$ and $\alpha_m$ that need to be defined and calibrated experimentally. These two variables are defined as:

$$\alpha_m\left(\sqrt{\bar{z}_m}\right) = \begin{cases} 0, if \sqrt{\bar{z}_m} \leq \sigma_{\alpha m0} \\ a_{\alpha m1}\left(\sqrt{\bar{z}_m} - \sigma_{\alpha m0}\right)^2, if \sigma_{\alpha m0} < \sqrt{\bar{z}_m} < \sigma_{\alpha m1} \\ a_{\alpha m1}(\sigma_{\alpha m1} - \sigma_{\alpha m0})^2, if \sqrt{\bar{z}_m} > \sigma_{\alpha m1} \end{cases} \qquad (A.13)$$

$$\alpha_f\left(\sqrt{\bar{z}_f}\right) = \begin{cases} 0, if \sqrt{\bar{z}_f} \leq \sigma_{\alpha f0} \\ a_{\alpha f1}\left(\sqrt{\bar{z}_f} - \sigma_{\alpha f0}\right)^2, if \sigma_{\alpha f0} < \sqrt{\bar{z}_f} < \sigma_{\alpha f1} \\ a_{\alpha f1}(\sigma_{\alpha f1} - \sigma_{\alpha f0})^2 + a_{\alpha f2}\left(\sqrt{\bar{z}_f} - \sigma_{\alpha f1}\right), if \sigma_{\alpha f1} < \sqrt{\bar{z}_f} < \sigma_{\alpha f2} \\ composite\, fracture\, when \sqrt{\bar{z}_f} > \sigma_{\alpha f2} \end{cases} \quad (A.14)$$

where $a_{\alpha m1}, \sigma_{\alpha m0}, \sigma_{\alpha m1}, a_{\alpha f1}, \sigma_{\alpha f0}$ are the model constants that need to be determined.

### 8.1.3 INELASTIC BEHAVIOR

To model the inelastic part of the CMC behavior, the assumption of the additive decomposition of the total strain into elastic and residual parts ($\boldsymbol{D}^e$ and $\boldsymbol{D}^r$, respectively) is assumed, that is,

$$\boldsymbol{D} = \boldsymbol{D}^e + \boldsymbol{D}^r \quad (A.15)$$

The evolution of the residual strain $\boldsymbol{D}^r$ is modeled using two uncoupled classical formulations of associated plasticity with isotropic hardening (one in each tow direction):

$$\boldsymbol{D}^r = \boldsymbol{D}^{r_1} + \boldsymbol{D}^{r_2} \quad (A.16)$$

For instance, regarding the longitudinal tow, the following effective and equivalent stresses are considered:

$$\begin{cases} \overline{\boldsymbol{\sigma}_1} = P\, \mathbb{A}\mathbb{A}_0^{-1}\boldsymbol{\sigma}^+ \\ \overline{\sigma}_1^{eq} = \sqrt{tr(\overline{\boldsymbol{\sigma}}_1^2)} \end{cases} \quad (A.17)$$

with $\boldsymbol{P} = \begin{matrix} 1 & 0 & 0 \\ 0 & 0 & 0 \\ 0 & 0 & \beta \end{matrix}$

where β is a model parameter defining the influence of shear on inelasticity. The yield surface is defined simply by the characteristic function:

$$f = \sigma_1^{eq} - R(r_1) - R_0 \qquad (A.18)$$

where $r_1$ is the cumulative residual strain and $R$ is the classical hardening function, which must be calibrated experimentally. Thus, the associated plasticity principle defines the following evolution law:

$$\begin{cases} \dot{\bar{D}}^{r_1} = \dot{\mu}\dfrac{\partial f}{\partial \sigma} = \dot{\mu}\dfrac{\overline{\sigma}_1}{\sqrt{tr(\overline{\sigma}_1^2)}} \\ \dot{r}_1 = -\dot{\mu}\dfrac{\partial f}{\partial R} = \dot{\mu} \end{cases} \qquad (A.19)$$

where $\dot{\bar{D}}^{r_1}$ is the effective residual strain rate and $\dot{\mu}$ is the residual multiplier rate, which can be calculated through the consistency condition. Finally, to keep the dissipation constant, the actual residual strain rate is given by:

$$\dot{D}^{r_1} = P\,\mathbb{A}\,\mathbb{A}_0^{-1}\dot{\bar{D}}^{r_1} \qquad (A.20)$$

### 8.1.4 Fracture Criterion

While the model described above can predict the microscopic damage which develops within the material, it cannot predict macroscopic fracture. Adoption of an ad hoc fracture criterion presumes that the composite fails when the effective damage parameter $\bar{z}_{f_1}$ exceeds some fixed value, that is,

$$\sqrt{\bar{z}_{f_n}} > \sigma_{\alpha f_{2n}}, \qquad n = \{1,2\} \qquad (A.21)$$

where $\sigma_{\alpha f_{2n}}$ is a fixed constant and $n$ represents the direction of the fiber.

## 9.0 TABLE OF FIGURES

## 10.0 TABLE OF TABLES

**11.0 TABLE OF CONTENTS**